\documentclass[
]{ceurart}

\usepackage{listings}
\usepackage{latexsym}
\usepackage{amsmath}
\usepackage{subcaption}
\usepackage{microtype}
\usepackage{graphicx}
\usepackage{acronym}
\usepackage{multicol}
\usepackage{multirow}
\usepackage{booktabs}
\usepackage{xcolor}
\usepackage{enumitem}
\usepackage{balance}

\definecolor{darkgreen}{RGB}{0,120,0}
\definecolor{green}{RGB}{0,180,0}
\definecolor{darkred}{RGB}{210, 0, 0}
\definecolor{red}{RGB}{250, 0, 0}
\definecolor{gray}{RGB}{60, 60, 60}
\definecolor{lightgray}{RGB}{180, 180, 180}
\definecolor{blue}{RGB}{0, 0, 200}
\definecolor{darkblue}{RGB}{0, 0, 150}

\newcommand{\gain}[1]{\textcolor{darkgreen}{\small{$\uparrow$\textbf{#1}\%}}}
\newcommand{\drop}[1]{\textcolor{darkred}{\small{$\downarrow$\textbf{#1\%}}}}

\acrodef{LLM}{Large Language Model}
\acrodef{PRP}{Pairwise Ranking Prompting}
\acrodef{PRD}{Pairwise Ranking Distillation}
\acrodef{RG}{Relevance Generation}

\begin{document}

\copyrightyear{2025}
\copyrightclause{Copyright for this paper by its authors. Use permitted under Creative Commons License Attribution 4.0 International (CC BY 4.0).}
\conference{ReNeuIR 2025 (at SIGIR 2025) -- 4th Workshop on Reaching Efficiency in Neural Information Retrieval, July 17, 2025, Padua, Italy}

\title{Harnessing Pairwise Ranking Prompting Through Sample-Efficient Ranking Distillation}

\author[1]{Junru Wu}[%
email=junru@google.com,
]
\cormark[1]

\author[1]{Le Yan}[%
email=lyyanle@google.com,
]

\author[1]{Zhen Qin}[%
email=zhenqin@google.com,
]

\author[1]{Honglei Zhuang}[%
email=hlz@google.com,
]

\author[2]{Paul Suganthan~G.~C.}[%
email=pachristopher@google.com,
]

\author[1]{Tianqi Liu}[%
email=tianqiliu@google.com,
]

\author{Zhe Dong}[%
email=hoogendong@gmail.com,
]
\fnmark[2]

\author[1]{Xuanhui Wang}[%
email=xuanhui@google.com,
]

\author[3]{Harrie Oosterhuis}[%
email=harrie.oosterhuis@ru.nl,
]
\fnmark[1]

\address[1]{Google DeepMind, New York City/Mountain View/Seattle, NY/CA/WA, USA}
\address[2]{Google, Zurich, Switzerland}
\address[3]{Radboud University, Nijmegen, The Netherlands}

\cortext[1]{Corresponding author.}
\fntext[1]{Work done while Harrie was at Google DeepMind.}
\fntext[2]{Work done while Zhe was at Google.}

\begin{abstract}
While \ac{PRP} with \acp{LLM} is one of the most effective zero-shot document ranking methods, it has a quadratic computational complexity with respect to the number of documents to be ranked, as it requires an enumeration over all possible document pairs.
Consequently, the outstanding ranking performance of PRP has remained unreachable for most real-world ranking applications.

In this work, we propose to harness the effectiveness of \ac{PRP} through pairwise distillation.
Specifically, we distill a pointwise student ranker from pairwise teacher labels generated by \ac{PRP}, resulting in an efficient student model that retains the performance of \ac{PRP} with substantially lower computational costs.
Furthermore, we find that the distillation process can be made sample-efficient: with only $2\%$ of pairs, we are able to obtain the same performance as using all pairs for teacher labels.
Thus, our novel approach provides a solution to harness the ranking performance of PRP without incurring high computational costs during both distillation and serving.
\end{abstract}


\maketitle
\acresetall
\section{Introduction}

\acp{LLM} have demonstrated incredible zero-shot performance across a wide range of natural language tasks including question answering, text summarization and ranking~\cite{openai2023gpt4, team2023gemini}.
However, \acp{LLM} need to be prompted correctly to be the most effective.
For document ranking, \ac{PRP} \cite{qin-etal-2024-large} achieves the state-of-the-art ranking performance, even with moderate-sized open-sourced \acp{LLM}.
In contrast, pointwise prompting also known as \ac{RG}~\cite{liang2022holistic, baidugpt} performs much more poorly and require large-sized LLMs.
The crucial difference between \ac{RG} and \ac{PRP} is that RG asks for the relevance of an individual document, i.e., ``\emph{How relevant is document A?}", whereas \ac{PRP} asks for relative relevance differences, i.e., ``\emph{Is document A more relevant than document B?}"
However, while being much more effective due to its pairwise approach, \ac{PRP} requires the prompting and aggregation for all document pairs to get a final ranking, resulting in a quadratic computational complexity: $O({N}^{2})$.
Consequently, the computational costs of PRP make it infeasible for any practical ranking setting where responsiveness is important.
In stark contrast, \ac{RG} only requires a single prompt per document, resulting in a more practical linear complexity: $O({N})$, but also substantially lower ranking performance.

In this paper, we aim to marry the strengths of both, i.e., inherit the \emph{effectiveness} of \ac{PRP} and the \emph{efficiency} of \ac{RG}.
To this end, we propose the novel \acfi{PRD} which distills the ranking ability from a pairwise \ac{LLM} rater to a more efficient pointwise \ac{LLM} ranker, without requiring an enumeration over all document pairs.
In summary, our main contributions are:
\begin{itemize}[leftmargin=*]
    \item We propose the novel \acfi{PRD} method to distill pairwise LLM ranking raters into pointwise student rankers. The student rankers can maintain the same ranking performance as PRP and significantly outperform the student rankers distilled from pointwise LLM ranking raters.
    \item We find that \ac{PRD} is sample-efficient and can amortize the quadratic complexity of \ac{PRP} during distillation. With only $2\%$ of pairs, it achieves performance comparable to $100\%$ of pairs.
    \item We design a novel ranking-aware sampling scheme that takes the order of documents in an early-stage ranking into consideration.
    This scheme needs less than $2\%$ of pairs for distillation without sacrificing performance, saving more training time.
\end{itemize}

\begin{figure} %
\centering
\includegraphics[width=0.92\textwidth]{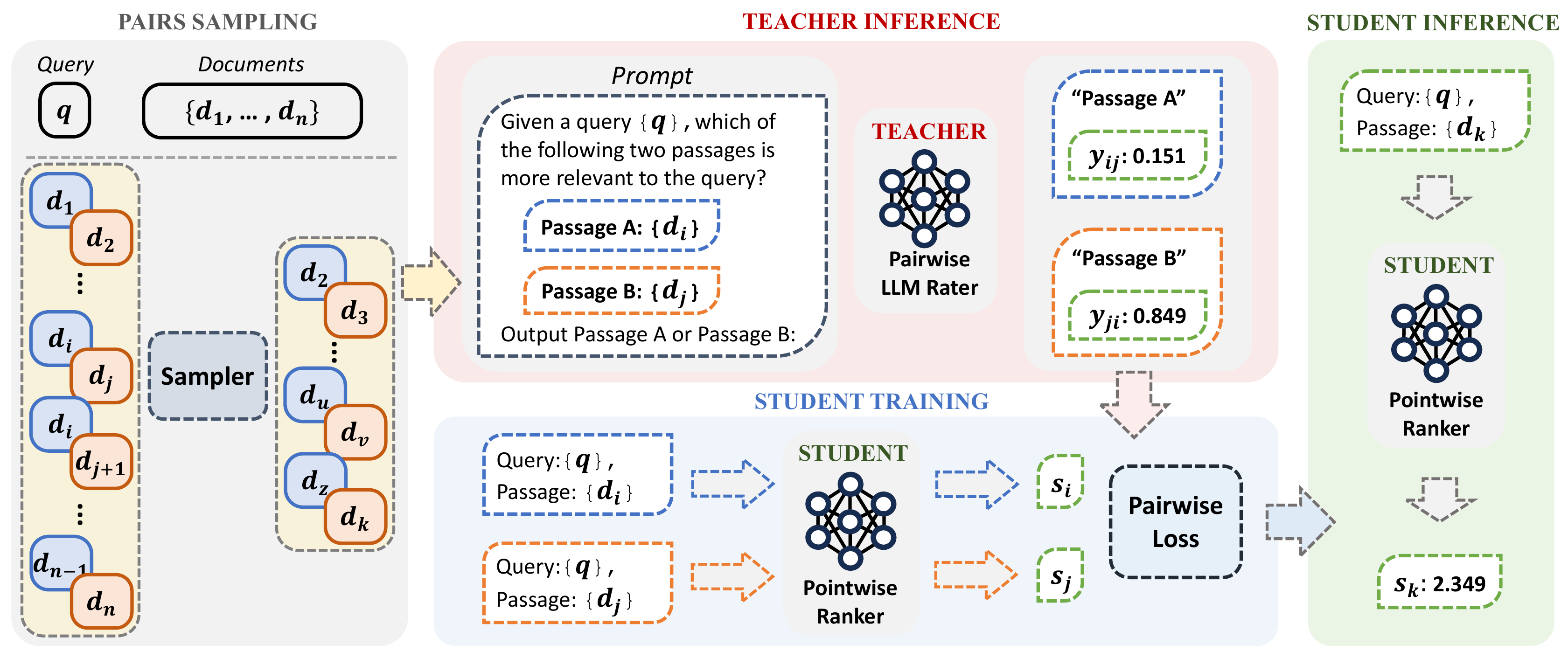}
\caption{
Visualization of the pipeline of our proposed Pairwise Ranking Distillation (PRD) method.
It can be divided into four stages (1) Pairs Sampling (2) Teacher Inference (3) Student Training and (4) Student Inference.
}
\label{fig:pipeline}
\end{figure}

\section{Related Work}

\paragraph{\textbf{Ranking with LLMs.}}
There have been many existing efforts to use LLM as zero-shot rankers in ranking tasks, this includes pointwise~\cite{baidugpt, zhuang2023beyond}, pairwise~\cite{qin-etal-2024-large} and listwise~\cite{pradeep2023rankvicuna} approaches.
By prompting LLMs differently, they are able to achieve different trade-offs between efficiency and effectiveness.
Among these zero-shot LLM rankers, listwise ranker are the most efficient but also the least effective, due to the limited  long-context capabilities of LLMs.
Conversely, pairwise rankers (i.e., through PRP) are the most effective but least efficient due to their quadratic complexity from enumerating all pairs.
On the other hand, pointwise rankers (i.e., through RG) are the most widely-used in real-world deployment due to their scalability.
The pointwise rankers effectiveness and efficiency falls between those of pairwise and listwise rankers, thus a natural extension could attempt to distill the strength of pairwise PRP approach, into the more scalable and efficient pointwise ranker. %

\paragraph{\textbf{Ranking Distillation.}}
Several existing works focus on distilling large neural rankers into more compacts models~\cite{qin2023rd} through Knowledge Distillation (KD)~\cite{hinton2015distilling}.
\citet{tang2018ranking} use a pointwise sigmoid cross-entropy loss and only optimize on positive examples from top-k ranked samples.
The later Rankdistil~\cite{reddi2021rankdistil} performs listwise optimization that penalizes large scores for items ranked low by the teacher to preserve its top-k ordering.

\paragraph{\textbf{Pairwise Distillation.}}
Instruction Distillation~\cite{sun2023instruction} distills the pairwise ranking capability of LLMs into a simpler but more efficient pointwise ranking model. More recently, PAIRDISTILL \cite{huang2024pairdistill} also aim to distill pairwise rankings, but into a dense retrieval models, which is usually a bi-encoder model.
Different from our \ac{PRD} approach, both \citet{huang2024pairdistill} use pairwise teacher supervision on top of a pointwise loss, and both \citet{huang2024pairdistill} and \citet{sun2023instruction} only use a naive random strategy for sampling pairs.
Unfortunately, none of these methods have been compared with any pointwise distillation baselines, therefore, it remains unclear how much can be gained from distilling a pairwise LLM teacher over a pointwise one.

\section{Problem Formulation and Background}
In the relevance ranking setting, the goal is to rank a candidate set of documents, $D=\{d_{1},...,d_{n}\}$, for a given query.
We use $Y=\{y_{1},...,y_{n}\}$ to denote their ground-truth labels, where $y_i$ is the ground-truth relevance of document $d_i$ to the query $q$.
The optimal ranking, $R=(r_{1},...,r_{n})$, orders the documents according to their ground-truth relevance in descending order, where documents with higher relevance should have smaller indices in $R$.
The problem is that the ground-truth labels are not available during inference, and thus the ranking has to be based on other features of the documents.

\label{sec:pointwise_llm_rater}
\paragraph{\textbf{Pointwise LLM Ranker.}} A straight-forward solution is to let an LLM directly predict the relevance labels, and subsequently, produce a ranking by sorting according to the predicted relevance.
This approach of predicting the relevance per individual query-document pair is named the \emph{pointwise} approach.
A specific example is the \acf{RG} approach proposed by \citet{liang2022holistic} and \citet{baidugpt}. Here an LLM prompted with $\mathcal{I}_\text{RG}$=“\emph{Does the passage \{document\} answer the query \{query\}? Output Yes or No:~}” for each query-document pair.
Then, the probability of the LLM generating the generated tokens ``\emph{Yes}'' or ``\emph{No}'' are normalized into query-document relevance scores following:
\begin{equation}
s^{\text{RG}}_{q,d} = \frac{p(\text{``\emph{Yes}''} \mid \mathcal{I}_{RG}(q,d))}{p(\text{``\emph{Yes}''}\,|\, \mathcal{I}_{RG}(q,d)) \!+\! p(\text{``\emph{No}''}\,|\, \mathcal{I}_{RG}(q,d))}.
\label{eq:rg_score}
\end{equation}
In practice, for instruction-tuned LLMs, the sum of the two likelihoods is often very close to one, so one can directly use the unnormalized ``\emph{Yes}'' probability.
The resulted ranking $R$ is obtained by sorting according to the relevance scores.
\label{sec:pairwise_llm_rater}
\paragraph{\textbf{Pairwise LLM Ranker.}} \acp{LLM} have been found to be better at ranking with a pairwise prompting approach~\cite{qin-etal-2024-large}, where the LLM is asked to judge a pair of documents and query on whether one document is more relevant to the query than the other.
\ac{PRP} uses the following pairwise prompt: $\mathcal{I}_{\text{PRP}}$=“\emph{Which of the following two passages is more relevant to the query \{query\}? Passage A: \{document A\}; Passage B: \{document B\}; Output Passage A or Passage B:~}”.
For each query-documents triplet $(q, d_{i}, d_{j})$, one of the following comparison results is obtained:
\begin{equation}
c_{ij} = \begin{cases} 
1, & \text{if } f(\mathcal{I}_{PRP}(q,d_i,d_j)) = d_i, \\
0, & \text{if } f(\mathcal{I}_{PRP}(q,d_i,d_j)) = d_j, \\
0.5, & \text{otherwise.}
\end{cases}
\label{eq:prp_score}
\end{equation}
where $f(\cdot)$ denotes the LLM output given the input prompt.
To obtain the final ranking, first an aggregation is performed over all pair comparisons with the following scoring scheme \cite{qin-etal-2024-large}:
\begin{equation}
s_{q,d}^{\text{PRP}} = \sum_{j \neq i} \left[c_{ij} + (1 - c_{ji})\right].
\label{eq:aggregate}
\end{equation}
Then a ranking is created by sorting according to the scores.
Because LLMs can be sensitive to the order of $i$ and $j$ in the prompt, comparisons in both directions are considered (both $c_{ij}$ and $c_{ji}$).

\section{Method: \acl{PRD}}
\label{sec:method}

Figure~\ref{fig:pipeline} shows the full pipeline of our proposed Pairwise Ranking Distillation (PRD) method, which efficiently leverages pairwise scores of LLM teachers to distill pointwise student rankers.
Our PRD consists of four stages: (1) Pairs Sampling, (2) Teacher Inference, (3) Student Training, and (4) Student Inference.
In this section, we further detail the distillation objective and the pair sampling of PRD.

\subsection{Distillation Objective}
\label{sec:distill_objective}
Given the teacher pseudo-label pairs, $y_{ij}$ and $y_{ji}$, we consider the Pairwise Logistic Ranking Loss~\citep{burges2005learning, burges2010ranknet}, which aims to encourage models to assign higher scores to positive instances than negative instances.
This can be formulated as the following loss function:
\begin{equation}
\label{eq:pair_logistic}
\mathcal{L} = \sum_{\langle i, j\rangle} \mathbf{1}_{y_{ij} < y_{ji}} \log\left(1 + \exp\left(s_i - s_j\right)\right),
\end{equation}
where $s_i$ and $s_j$ are the student predictions on query-documents pairs $(q, d_{i})$ and $(q, d_{j})$, respectively.
The summation is over all sampled pairs, the sampling procedure is described in Section~\ref{sec:pair_sampling}.

\subsection{Efficient Pair Sampling}
\label{sec:pair_sampling}
The distillation objective defined in Section~\ref{sec:distill_objective} involves a summation over a set of sampled query-document pairs.
This set could include all possible pairs, similar to the original Pairwise Ranking Prompting \cite{qin-etal-2024-large} approach resulting in ${N}^{2} - N$ pairs, which becomes prohibitively expensive for larger numbers of documents $N$.
However, unlike the setting in \citet{qin-etal-2024-large}, we have access to the initial ranking from relatively cheap (pointwise) rankers, such as a cheaper pointwise LLM ranker. 
To avoid the quadratic costs, we propose a number of novel efficient sampling strategies based on the initial ranking to drastically reduce the cost of invoking pairwise LLM rankers.

\paragraph{Random Sampling.}
The most straightforward method is uniform random sampling.
For all the possible pairs $(d_i, d_j)$ where $i \neq j$, we uniformly randomly sample $k$ pairs and obtain their pairwise comparison prediction from the LLM pairwise ranker.

\paragraph{Reciprocal Rank Weighted (RR).} %
To capture the most prominent ranking metrics, e.g., the Mean Reciprocal Rank~\cite{Craswell2009}.
Accordingly, we propose a sampling strategy that is more likely to sample pairs that would be helpful to correct top-ranked items.
This strategy assumes access to an initial ranking obtained by a pointwise ranking method such as a pointwise LLM ranker, where the ranked position of each $d_i$ is denoted by $r_i \in \{1, 2, \ldots, n\}$.
For all the possible pairs $(d_i, d_j)$ where $i \neq j$, we can assign an unnormalized weight $w_{ij}$ as the reciprocal rank of the first document $d_i$:
$w_{ij} = \frac{1}{r_i}.$
Then a fixed number $k$ of pairs $(d_i, d_j)$ are sampled, without replacement, from all the possible pairs with a probability proportional to the weights $w_{ij}$ of each pair, which gives pairs with a high-ranked document in the initial ranking a higher chance to be re-examined by the more accurate pairwise LLM ranker.

\paragraph{Reciprocal Rank Sum Weighted (RRSum).}
As a natural extension of the RR method, we propose a strategy that considers the reciprocal rank of both items by taking the average reciprocal rank.
Analogous to the RR strategy, the RRSum weights for each pair $(d_i, d_j)$ are defined as:
$w_{ij} = \frac{1}{2}\Big(\frac{1}{r_i} + \frac{1}{r_j}\Big).$
Correspondingly, RRSum also samples $k$ pairs, without replacement, with a probability proportional to its unnormalized weights.

\paragraph{Reciprocal Rank Diff Weighted (RRDiff).}
Another possible intuition is to sample pairs with the largest potential impact on the overall ranking quality relative to the initial ranking quality.
We note that several learning-to-rank algorithms follow similar intuitions~\cite{burges2010ranknet,burges2011learning,wang2018lambdaloss}.
Accordingly, we propose RRDiff which assigns weights to document pairs based on the \emph{absolute difference} between their reciprocal ranks. 
We define these weights for each pair $(d_i, d_j)$ as:
$w_{ij}=\Big|\frac{1}{r_{i}}-\frac{1}{r_{j}}\Big|$.
Again, $k$ pairs are sampled, without replacement, with probabilities proportional to the defined weights.
Unlike RRSum, RRDiff can give pairs of documents that are both ranked higher in the initial ranking lower weights than pairs of documents where one document is ranked much higher than the other.

\label{sec:student_inference}

\begin{table}
\caption{Results on TREC-DL datasets with BM25 retrieved top-100 passages. Best performing system in each model variant is marked in bold. \emph{Agg.}\ denotes whether to do full-pair aggregation according to Eq.~\ref{eq:aggregate}. We used the random sampling strategy for sampling the subsets of training pairs.}
\label{tbl:trecdl}
\resizebox{0.999\textwidth}{!}{%
\setlength{\tabcolsep}{2pt}%
\begin{tabular}{l|c|c|c|cccccccc|cc}
\toprule
\multirow{2}{*}{\textbf{\begin{tabular}{@{}c@{}}LLM \\ Rater\end{tabular}}} & \multirow{2}{*}{\textbf{Model}} & \multirow{2}{*}{\textbf{Agg.}} & \multirow{2}{*}{\textbf{\#Pairs}} & \multicolumn{2}{|c|}{\textbf{TREC-DL2019}} & \multicolumn{2}{|c|}{\textbf{TREC-DL2020}} & \multicolumn{2}{|c|}{\textbf{TREC-DL2021}} & \multicolumn{2}{|c|}{\textbf{TREC-DL2022}} & \multicolumn{2}{|c}{\textbf{Average}} \\
& & & & OPA & nDCG@10 & OPA & nDCG@10 & OPA & nDCG@10 & OPA & nDCG@10 & OPA & nDCG@10 \\
\midrule
- & \multirow{1}{*}{BM25} & - & - & 59.76 & 46.22 & 70.57 & 55.20 & 57.18 & 34.68 & 45.69 & 27.41 & 58.30 & 40.88 \\
\midrule
\multicolumn{14}{c}{\textbf{Teacher}} \\
\midrule
\midrule
Pointwise & \multirow{2}{*}{\begin{tabular}{@{}c@{}}PaLM 2 \\ L\end{tabular}} & - & - & 83.70 & 74.13 & 81.76 & 69.73 & 89.90 & 81.20 & 82.98 & 69.59 & 84.59 & 73.66 \\
Pairwise & & - & - & 88.33 & 79.05 & 82.83 & 70.13 & 91.53 & 83.41 & 85.88 & 72.06 & 87.14 & 76.16 \\
\midrule
\multicolumn{14}{c}{\textbf{Student}} \\
\midrule
\midrule
\multirow{1}{*}{Pointwise} 
& \multirow{4}{*}{\begin{tabular}{@{}c@{}}Gemma1 \\ 2B\end{tabular}} & - & - & 75.60 & 68.39 & 75.46 &	67.22 & 74.60 & 60.43 & 71.42 & 50.15 & 74.27 & 61.55 \\
\cline{1-1}\cline{3-14}
\multirow{3}{*}{\makecell{Pairwise\\(PRD)}} & & Yes & $100\%$ & 87.78\gain{12.1} & 79.45\gain{11.0} & 82.36\gain{6.9} & 72.86\gain{5.6} & \textbf{88.16\gain{13.5}} & \textbf{76.33\gain{15.9}} & \textbf{82.55\gain{11.1}} & \textbf{71.65\gain{21.5}} & 85.21\gain{10.9} & \textbf{75.07\gain{13.5}} \\
& & No & $100\%$ & \textbf{88.18\gain{12.5}} & \textbf{80.11\gain{11.7}} & \textbf{83.62\gain{8.1}} & \textbf{73.58\gain{6.3}} & 87.81\gain{13.2} & 74.30\gain{13.8} & 82.17\gain{10.7} & 68.26\gain{18.1} & \textbf{85.45\gain{11.1}} & 74.06\gain{12.5} \\
& & No & $2\%$ & 84.09\gain{8.4} & 70.59\gain{2.2} & 82.29\gain{6.8} & 73.36\gain{6.3} & 87.65\gain{13.0} & 75.03\gain{14.6} & 82.25\gain{10.8} & 68.39\gain{18.2} & 84.07\gain{9.8} & 71.85\gain{10.3} \\
\midrule
\multirow{1}{*}{Pointwise} & \multirow{4}{*}{\begin{tabular}{@{}c@{}}Gemma2 \\ 2B\end{tabular}} & - & - & 80.83 & 67.37 & 84.40 & \textbf{74.56} & 86.17 & 74.87 & 77.00 & 61.08 & 82.10 & 69.47 \\
\cline{1-1}\cline{3-14}
\multirow{3}{*}{\makecell{Pairwise\\(PRD)}} & & Yes & $100\%$ & \textbf{88.24\gain{7.4}} & 79.46\gain{12.1} & 83.10\drop{1.3} & 73.06\drop{1.5} & 88.77\gain{2.6} & \textbf{77.99\gain{3.1}} & \textbf{84.62\gain{7.6}} & \textbf{75.24\gain{14.1}} & 86.18\gain{4.0} & 76.44\gain{6.9} \\
& & No & $100\%$ & 88.19\gain{7.3} & \textbf{83.39\gain{16.0}} & 83.48\drop{0.92} & 72.59\drop{1.97} & \textbf{89.03\gain{2.8}} & 77.68\gain{2.8} & 84.09\gain{7.1} & 73.75\gain{12.6} & \textbf{86.20\gain{4.1}} & \textbf{76.85\gain{7.4}} \\
& & No & $2\%$ & 85.06\gain{4.2} & 75.16\gain{7.8} & \textbf{85.38\gain{0.9}} & 74.46\drop{0.1} & 87.80\gain{1.6} & 75.33\gain{0.46} & 84.00\gain{7.0} & 69.52\gain{8.4} & 85.56\gain{3.4} & 73.62\gain{4.1} \\
\midrule
\multirow{1}{*}{Pointwise} & \multirow{4}{*}{\begin{tabular}{@{}c@{}}Gemma1 \\ 7B\end{tabular}} & - & - & 82.78 & 69.98 & 77.13 & 70.21 & 82.26 & 72.03 & 77.42 & 59.44	& 79.90 & 67.92 \\
\cline{1-1}\cline{3-14}
\multirow{3}{*}{\makecell{Pairwise\\(PRD)}} & & Yes & $100\%$ & 87.70\gain{4.9} & \textbf{80.54\gain{10.5}} & 80.68\gain{3.5} & 68.50\drop{1.7} & \textbf{83.76\gain{1.5}} & \textbf{72.37\gain{0.3}} & \textbf{81.79\gain{4.3}} & \textbf{69.26\gain{9.8}} & \textbf{83.48\gain{3.5}} & \textbf{72.67\gain{4.7}} \\ %
 & & No & $100\%$ & \textbf{87.85\gain{5.1}} & 78.51\gain{8.5} & 80.98\gain{3.8} & 70.95\gain{0.7} & 83.35\gain{1.1} & 66.79\drop{5.2} & 80.55\gain{3.1} & 65.55\gain{6.1} & 83.19\gain{3.3} & 70.45\gain{2.5} \\
& & No & $2\%$ & 86.23\gain{3.4} & 79.45\gain{9.4} & \textbf{81.30\gain{4.2}} & \textbf{71.71\gain{1.5}} & 83.72\gain{1.4} & 70.55\drop{1.4} & 79.29\gain{1.8} & 63.32\gain{3.8} & 82.64\gain{2.7} & 71.26\gain{3.3} \\
\bottomrule
\end{tabular}}
\end{table}

\section{Experiments and Results}
\subsection{Datasets.}
The TREC (Text REtrieval Conference) dataset refers to a collection of datasets, that offers a common ground for evaluating different information retrieval systems in IR Research~\citep{voorhees2005trec}. TREC-DL (Deep Learning Track) was specifically design for evaluating the task of passage re-ranking~\citep{craswell2021trec, craswell2020overview}.
TREC-DL 2019, 2020, 2021, 2022 subsets contains 43, 54, 53, 76 queries, respectively. For each query, we used the initial ranking of 100 passages from a IR system (i.e., BM25 score), then re-rank the passages based on their relevance of an answer to the question. 
We split the dataset by queries in a 7:1:2 split as training / validation / testing sets for supervised fine-tuning.

\subsection{Implementation Details.}
\paragraph{\textbf{Teacher Model.}} We use instruction-tuned PaLM 2-L \cite{anil2023palm} as our pairwise LLM rater teacher, due to it being proven to be particular effective in the pairwise prompting setting due to its large model capacity.
Similar to Pairwise Prompt Ranking \cite{qin-etal-2024-large}, we used the \emph{scoring LLMs APIs} to avoid potential issues with the generation API as it is prone to generate undesired outputs.
Specifically, we used the same pairwise prompt as \citet{qin-etal-2024-large}, and used the log probability of token ``\emph{Passage A}'' and ``\emph{Passage B}'' as the output of the pairwise LLM rater teacher, as discussed in detail in Section~\ref{sec:pairwise_llm_rater}.
\paragraph{\textbf{Student Model.}} Decoder-only architecture are widely used in modern Large Language Models (LLMs) such as GPT \cite{achiam2023gpt}, Llama \cite{dubey2024llama3, touvron2023llama2} and Gemma \cite{team2024gemma, team2024gemma2}. However, such decoder-only architecture generate tokens autoregressively, and are designed for generative tasks only, such as question answering, summarization, etc.
As a result, decoder-only architectures are generally not suitable for discriminative tasks that require single or multiple scalars as output.
In light of this, we convert a decoder-only model to become encoder-only following Gemma Encoder\cite{suganthan2025adapting}, such that the model can output a single scalar for ranking regression. 
We use Gemma1-2B, Gemma1-7B, and Gemma2-2B from the instruction-tuned Gemma1~\cite{team2024gemma} and Gemma2~\cite{team2024gemma2} model families as our model backbones and initialization checkpoints.

\begin{figure} %
\centering
\begin{subfigure}{0.5\textwidth}
\centering
\includegraphics[width=\textwidth]{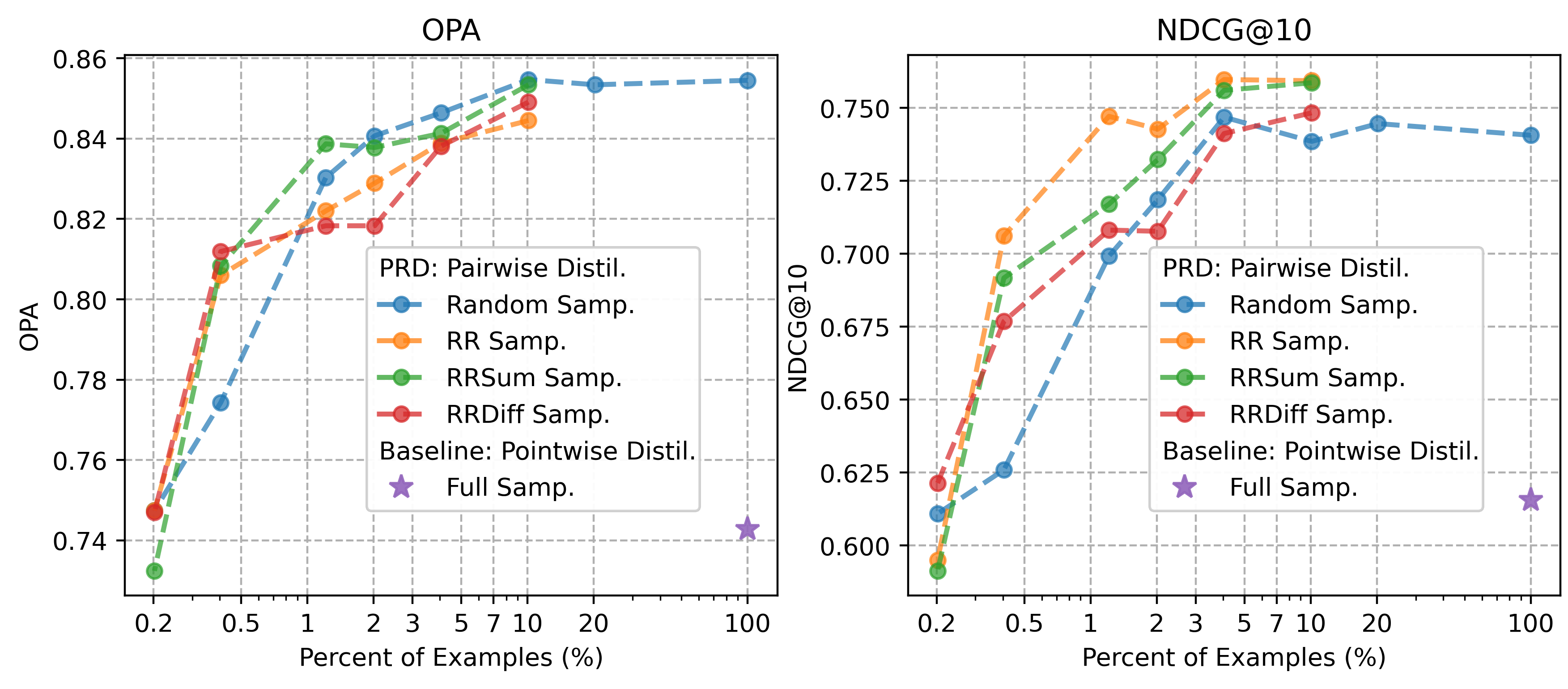}
\caption{OPA over \% of training examples (sampled pairs).}
\label{fig:ablation_sampling_trecdl_percent}
\end{subfigure}\hfill
\begin{subfigure}{0.5\textwidth}
\centering
\includegraphics[width=\textwidth]{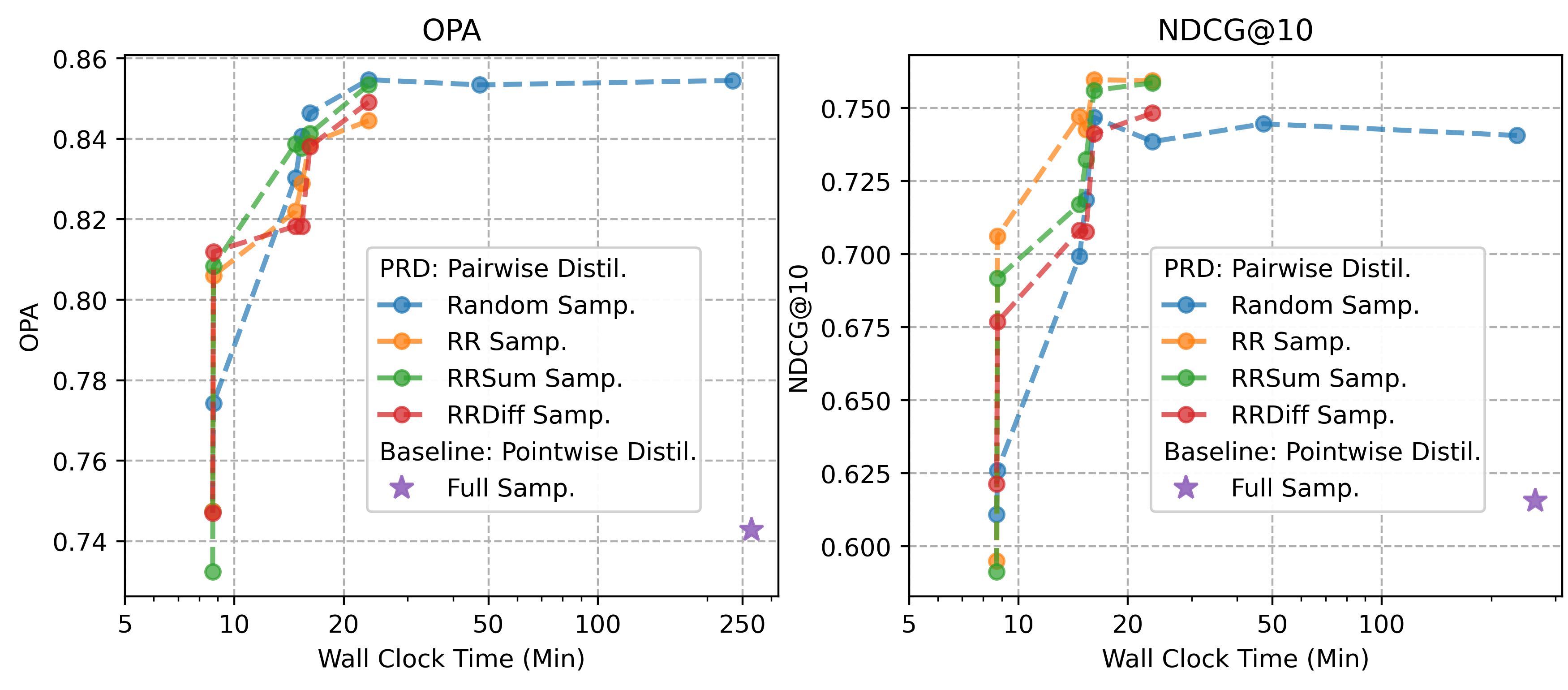} %
\caption{OPA over wall-clock time.}
\label{fig:ablation_sampling_trecdl_wallclock}
\end{subfigure}
\caption{Comparison of different sampling strategies on TREC-DL with Gemma1-2B as student backbone.}
\label{fig:ablation_sampling_trecdl}
\end{figure}

\subsection{Experimental Results}
We compare our proposed Pairwise Ranking Distillation (PRD) with the following baselines: (a) BM25, which is an unsupervised method based on weighted term frequency; (b) Pointwise Teacher, which is a pointwise LLM rater as the teacher described in Section~\ref{sec:pointwise_llm_rater}; (c) Pairwise Teacher, which is the pairwise LLM rater as the teacher described in Section~\ref{sec:pairwise_llm_rater}; (d) Pointwise Student, which is the pointwise LLM rater as the teacher described in Section~\ref{sec:pointwise_llm_rater} to finetune an encoder-only student model;
(e) Pairwise Student w/ Aggregation, which is the pairwise LLM rater as the teacher described in Section~\ref{sec:pairwise_llm_rater} to finetune a encoder-only student model, we do full-pair aggregation following Eq.~\ref{eq:aggregate};
(f) Pairwise Student w/o Aggregation, which is our proposed Pairwise Ranking Distillation (PRD) approach.
We were unable to repeat these runs due to the high costs of training each student model.

Our experiments on TREC-DL represent passage re-ranking tasks; and our metrics are the pairwise ordered pair accuracy (OPA) and the listwise NDCG@10.
Overall, our PRD method achieves competitive performance;
we observe the following in Table~\ref{tbl:trecdl} and Figure~\ref{fig:ablation_sampling_trecdl_percent}:
\begin{itemize}[leftmargin=*]
\item PRD outperforms pointwise distillation in all cases by a significant margin (e.g., 3\%-13\% improvement in OPA and NDCG metric), regardless of the model backbone, and across all datasets
\item There is no meaningful difference between full-pair aggregation according to Eq.~\ref{eq:aggregate} and PRD sampling all the pairs independently.
\item By employing a simple yet effective random sampling scheme, PRD achieves comparable performance to full pairs with only $2\%$ of the pairs, demonstrating its high sample-efficiency and significant saving in training wall-clock time.
\item Furthermore, ranking-aware sampling techniques, in particular RR, further boosts the performance of ranking distillation when sampling less than $2\%$ of the pairs.
\end{itemize}

\section{Conclusion}
In this paper, we proposed Pairwise Ranking Distillation (PRD), a method that distills the ranking ability of a complex pairwise LLM rater into a more efficient pointwise LLM ranker.
Importantly, it does so without requiring an enumeration of all possible document pairs during training or inference, thus avoiding a quadratic complexity.

We found our proposed pairwise distillation scheme significantly outperforms approaches using the pointwise LLM ranking rater. Our results indicate that the distillation process can be performed sample-efficiently; by employing a simple yet effective random sampling scheme, we achieve comparable performance against full pairs, while using only 2\% of pairs and still surpassing the model distilled by a pointwise LLM rater.

Moreover, we also proposed several ranking-aware sampling schemes that samples pairs base on a first-stage retrieval ranking of documents.
With these schemes, we need less than $2\%$ of the all pairs to reach optimal performance after distillation.

\section*{Declaration on Generative AI}
The author(s) have not employed any Generative AI tools for writing the text in this paper.

\bibliography{references}

\appendix

\end{document}